\documentclass[amsfonts,aps, nofootinbib, showpacs, showkeys]{revtex4}

\usepackage{amsmath}
\usepackage{graphicx}% Include figure files
\usepackage{dcolumn}% Align table columns on decimal point
\usepackage{bm}% bold math

\begin{document}

\title[Evolutionary dynamics]{Evolutionary dynamics from a variational principle}

\author{Peter Klimek$^1$; Stefan Thurner$^{1,2,*}$; Rudolf Hanel$^1$;}

\affiliation{$^1$Complex Systems Research Group; Medical University of Vienna; 
W\"ahringer G\"urtel 18-20; A-1090; Austria; \\ $^2$Santa Fe Institute; 1399 Hyde Park Road; Santa Fe; NM 87501; USA}
\email{thurner@univie.ac.at}

\begin{abstract}
We demonstrate with a thought experiment that fitness-based population dynamical approaches to evolution are not able to make quantitative, falsifiable predictions about the long-term behavior of evolutionary systems.
A key characteristic of evolutionary systems is the ongoing endogenous production of new species. These novel entities change the conditions for already existing species. 
Even {\em Darwin's Demon}, a hypothetical entity with exact knowledge of the abundance of all species and their fitness functions at a given time, could not pre-state the impact of these novelties on established populations.
We argue that fitness is always {\it a posteriori} knowledge -- it measures but does not explain why a species has reproductive success or not. To overcome these conceptual limitations, a variational principle is proposed in a spin-model-like setup of evolutionary systems.
We derive a functional which is minimized under the most general evolutionary formulation of a dynamical system, i.e. evolutionary trajectories causally emerge as a minimization of a functional. 
This functional allows the derivation of analytic solutions of the asymptotic diversity for stochastic evolutionary systems within a mean-field approximation.
We test these approximations by numerical simulations of the corresponding model and find good agreement in the position of phase transitions in diversity curves. 
The model is further able to reproduce stylized facts of timeseries from several man-made and natural evolutionary systems.
Light will be thrown on how species and their fitness landscapes dynamically co-evolve.


%89.75.Fb   Self-organization complex systems
%05.65.+b   Self-organization statistical physics
%87.23.Cc   Ecology, population dynamics
%87.10.Ed   Integrodifferential equations in mathematical aspects of biological physics
%87.23.Kg   Evolution in biology

\end{abstract}

\maketitle

\section{Introduction}

Evolutionary dynamics appear in a multitude of different contexts. 
Evolution basically describes how sets of elements, such as  biological species, goods and services in an economy, groups of living beings, or chemical compounds, 
change over time. Examples are abundant in various areas. Chemical compounds react with other compounds to produce new chemicals. Integrated circuits performing specific computational tasks can be combined to create another circuit for a different computational task. Prey and predator may co-evolve by succinctly acquiring new traits and thereby develop into new species. In the following we will use {\em species} for elements in whatever context,  chemicals, goods, biological species, etc. 
The removal or addition of a single species in an evolutionary system may have dramatic consequences.
For example, in starfish removal experiments (e.g. Mukkaw Bay in Washington \cite{Paine69}) starfish are removed from an eco-system with the consequences that mussel populations explode and drive out most other species, while the urchin population destroys coral reefs. 
In 1904 English physicist John Ambrose Fleming accidentally manufactured the first vacuum tube which triggered a cascade of technological and economic co-evolutions and adaptations; in 2004 the semiconductor industry was a market of \$ 213 billion and enabled the generation of approximately \$ 1200 billion in electronic system businesses and \$ 5000 billion in turn in services which amounts to 10\% of world GDP \cite{SIA}.
Typically in  evolutionary systems species are endogenously added or removed from a large system of mutually influencing species.
Two species influence each other if the existence of one species has a positive or negative effect on  the change of abundance of the other. 
The possibilities for  interactions in evolutionary systems involve different natural, economic or social laws on a variety of  time or length scales.
The collective result of these ('microscopic') interactions between elements leads to ubiquitous well-known macro phenomena in evolutionary systems, 
such  as punctuated equilibria, booms of diversification, breakdowns and crashes, or seemingly 
unpredictable responses to external perturbations. 
Maybe one of the most exciting questions in natural sciences today is to understand if evolutionary dynamics can be understood by a common underlying principle and -- if yes -- how such a principle might look like. Such a principle must be general enough to capture the multitude of different phenomena, and at the same time must be in a form which can be applied easily to specific problems.

In the present understanding of evolution the concept of  {\em fitness} is of central importance. 
Usually the relative abundance of species (with respect to other species) is described by replicator equations (first-order differential equations) and variants such as Lotka-Volterra equations \cite{Volterra31, Crow70, Hofbauer98}.   
Their mutual influence is quantified by a rate specifying how the abundance of one species changes upon contact with another.
In biology this rate is called Malthusian fitness, in chemistry one refers to it as the reaction rate, in economics it is related to production functions. 
Similar proliferation rates could also be introduced for technological, financial or even historical contexts.  In the following we subsume them all under the term {\em fitness}.  

A distance between two species can be defined as the minimal number of evolutionary steps needed for one species to evolve into the other one 
\cite{Maynard70}  (in biology this distance is often the number of single-point mutations two species differ in).
In this way a metric is given on the space of all possible species.
A {\it fitness landscape} assigns to each point in this space (that is to each species) its reproductive success or fitness.
Evolution is sometimes pictured as an optimization problem where species evolve and co-evolve via adaptations toward peaks in this landscape \cite{Wright32}.
These peaks represent regions of high reproductive success, whereas valleys correspond to low fitnesses.
As these peaks and valleys are spread over larger regions of species space, the fitness landscape becomes less rugged.
If one species moves toward a peak it may change the fitness of other species which are then moving on this fitness landscape too \cite{kauffman_NK} -- one speaks of co-evolutionary cascades.  
The higher the degree of 'ruggedness' (i.e. the smaller the average distance between adjacent adaptive peaks or valleys), the higher the probability that these cascades of co-evolution last for long times.
With lower ruggedness, however, the probability increases that each species reaches an adaptive peak and the evolutionary dynamics comes to a halt.

The concept of fitness is limited however. To see this consider  the following thought experiment.
Suppose one -say a demon- would have exact knowledge about the abundance and fitness of each biological species in the universe. 
'Knowing the fitness of a species' means knowledge of the functional dependence of its proliferation rate on the entire current environment (i.e.  all other species).
The omniscient hypothetical entity in possession of this knowledge could be called {\em Darwin's Demon} for obvious reasons. 
The demon may be pictured as a super-biologist, able to measure each species' abundance as well as the dependence of its proliferation rate on each other species in each habitat.
That is, he knows the set of all existing species and can measure their associated fitness landscape to an arbitrary degree of exactness.
What can the demon {\it predict} about the future course of evolutionary events, such as biodiversity in 100 million years or the time to the next mass extinction event?
Surprisingly little, for the following reasons.
A key characteristic of evolutionary systems is its potential to generate innovations, i.e. new species. In biological systems this can happen  through mutations, in technological or economical ones through spontaneous ideas of an inventor, etc.
Once a new species is created it becomes part of the environment and thereby potentially changes the conditions for all already existing species and for those yet to arrive. To now assess the fitness of a new species one has to measure how it spreads in an environment it is now part of.
The demon has information related to a different environment, one which only existed {\em before} arrival of the new species.
Thus the demon may have an exact description of the {\it current} biosphere, but with the advent of each new species this description loses accuracy.
Fitness thus always encodes {\it a posteriori} knowledge, and can not be used to make falsifiable predictions.
It is not fruitful to predict future fitness of species from their present fitness.
Instead  one has to understand how species and their fitness landscapes co-construct each other, how they {\em co-evolve}.

To make headway in understanding the phenomenology of evolution, i.e. in identifying principles which guide evolutionary dynamics, a series of quantitative models have been suggested \cite{kauffman_NK, Bak93, solemanrubia, jainkrishna, newman, barthur, Dittrich07,Kauffman93, kardanoff}.
Here explicit assumptions are made about how new species come into being, how they interact with each other and under which conditions or under which selective forces they vanish. Each of these  models focuses on particular aspects of evolution. 
For example in Kauffman $NK$ models \cite{kauffman_NK} species are bit-strings with randomly assigned fitness values. 
Arthur \cite{barthur} focuses on technological evolution with integrated circuits as species whose fitness is examined by how well they execute certain computational tasks.  
Jain and Krishna \cite{jainkrishna} consider ecological systems and elucidate the interplay between interaction topology and survival of species.   
In most of these models some {\it ad hoc} assumptions about the  mechanisms have to be made.
In the model by Jain and Krishna, for example, species are actively removed and added to the system, innovations are externally enforced and not endogenously produced.
In \cite{barthur}  the output of randomly assembled  circuits  is compared to a prespecified list of desired computational tasks (such as bitwise addition).
In the $NK$ model evolutionary interactions are constrained to actions on bit-strings.
Although these assumptions are certainly reasonable in the specific contexts of their models, it is not at all clear whether conclusions derived on the basis of these assumptions are universally valid in different contexts.

To arrive at a general evolutionary description (without ad hoc specifications) one has to identify principles which are abstract enough to be applicable in each evolutionary context but which must be specific enough to make useful quantitative predictions.
To meet these requirements, evolution can be pictured as a three-step production/destruction process. 
{\bf Step 1:} New species come into being through recombination of already present species. 
That is, each species arises only under the condition that a given (and maybe not unique) set of other species or environmental factors exists.
For example, to assemble an MP3-player all parts --including software-- are needed. 
Sodium chloride can be produced by sodium hydroxide in solution with hydrochloric acid.
With a substantial degree of oversimplification one can say that apes in combination steppe formation give rise to mankind.
{\bf Step 2:} The new species becomes part of the system and can now be combined with other, already existing species.
One can legally download music for the MP3-player and listen to it, sodium chloride reacts with e.g. calcium carbonate in the Solvay process, mankind 
burns forests  to create fields for agriculture.
{\bf Step 3:} As a consequence, through this recombination yet new species may come into being and other already existing ones may vanish or be destroyed.
For example, MP3 currently drives CDs out of the market but can be combined with cell phones to give smartphones. 
Soda ash can be used to remove sulfur dioxide from flue gases in power stations
which might help to reduce the ongoing Holocene extinction event of biological species \cite{Pimm95} possibly influenced by the advent of mankind.
In previous work models incorporating these types of production and destruction processes have been shown to reproduce a wide range of evolutionary phenomena, including booms of diversification \cite{htk1}, breakdowns of diversity \cite{htk2} or punctuated equilibria \cite{thurner09}. 
Such processes further allow to understand stylized facts in time-series data on evolutionary systems, such as scale-free distributions of species lifetimes, the number of species per genus or the size of extinction events in fossil data \cite{klimek09}, or GDP and business failures in economic markets \cite{tkh09}.

In this work we propose a variational principle from which dynamics -- identical to the dynamics of the production / destruction processes described above -- can be derived.  
To this end we define  the {\it evolutionary potential} of a species. This function measures in how many productions and destructions a species would (no longer) take part if it would enter (be removed from) the system.
With this potential one obtains two formal representations of the system's dynamic:  
(i) The potential can be used to explicitly deduce a set of dynamical {\em update equations} of system diversity for production / destruction processes.
(ii) Using this evolutionary potential and a measure for ongoing productions and destructions one can derive a {\it balance function}.
The evolutionary process solving the dynamical update equations (i) always minimizes the balance function (ii).
The balance function further allows asymptotic solutions for the system diversity (mean-field approximation).
These analytic solutions are in good agreement with numerical simulations of the full model of productions and destructions. This is to a certain degree unexpected 
since the dynamics is dominated by strong and nonlinear interactions.

This description of evolutionary systems allows to understand how the set of existing species and their fitness landscapes co-construct each other from first principles,
as opposed to research strategies portrayed by Darwin's Demon, where snapshots of regions of fitness landscapes are empirically explored.
Accordingly the focus shifts from predicting {\em microscopic}  properties such as individual proliferation rates to estimating the occurrence of global, {\em macroscopic} events.

This work is structured as follows.  In section \ref{sec: GenFor} we develop a general framework for evolutionary systems via a variational principle. 
We discuss deterministic and stochastic implementations and  obtain asymptotic diversity solutions in a mean-field approximation. 
In section \ref{sec: MotivMod} we motivate  and define the choice of evolutionary interactions as production and destruction rules as in \cite{tkh09}. We 
treat the special cases of systems with only productive interactions in section \ref{sec: Constr} and the pure destructive case in section \ref{sec: Destr}. Then we discuss the full model of productions {\it and} destructions in section \ref{sec: CombDyn}. 
We discuss empirical relevance of this work  in section \ref{sec: EmpRel}   
 and turn to a conclude in  \ref{sec: Discuss}.

\section{General formulation of diversity dynamics}
\label{sec: GenFor}

\subsection{Dynamical systems}
The abundance of species $i$ is given by a binary state variable $\sigma_i(t) \in \{0,1\}$. If species $i$ exists at time $t$, $\sigma_i(t)=1$, otherwise $\sigma_i(t)=0$. The system can be populated by  $N$ species ($N$ arbitrarily large, even infinite).  A particular configuration of the system is characterized by the $N$-dimensional vector in phase space $\vec \sigma(t)=(\sigma_i(t)) \in \Gamma = \{0,1\}^{N}$. The system's {\it diversity} $D(t)$ is given by $D(t)=\frac1N \sum_i\sigma_i(t)$.

At each time, species $i$ may experience three scenarios, (i) annihilation $\sigma_i(t) =1 \to \sigma_i(t+1) = 0$, (ii) nothing $\sigma_i(t) = \sigma_i(t+1)$ or (iii) creation $\sigma_i(t) =0 \to \sigma_i(t+1) =1$. 
Suppose that there exists a function $f_i(\vec \sigma(t)) : \{0,1\}^{N} \to \mathbb R$ indicating which of the transitions (i)-(iii) takes place.
Specifically, let $f_i(\vec \sigma(t))$  indicate the following transitions 
\begin{eqnarray} \label{function1} 
\mathrm{(i) }\ & f_i(\vec \sigma(t))  <  0 & \Rightarrow   \sigma_i(t+1)=0 \nonumber  \\
\mathrm{(ii) }\ & f_i(\vec \sigma(t))   =  0 & \Rightarrow  \sigma_i(t+1)=\sigma_i(t) \\
\mathrm{(iii) }\ & f_i(\vec \sigma(t))   >  0 & \Rightarrow   \sigma_i(t+1)=1 \nonumber 
\end{eqnarray}
For (i) or (iii) a transition occurs if $\sigma_i(t)=1$ or $0$, respectively. That is, if $f_i(\vec \sigma(t)) \geq 0$ the system evolves according to 
\begin{equation}
\sigma_i(t+1)  =  \sigma_i(t)+\Delta \sigma_i(t) \qquad {\rm with} \qquad \Delta \sigma_i(t) =\mathrm{sgn} \left[(1-\sigma_i(t)) f_i(\vec \sigma(t)) \right].
\label{depp}
\end{equation}
$\Delta \sigma_i(t)$ can only be non-zero if $\sigma_i(t)=0$ {\it and} $f_i(\vec \sigma(t)) > 0$. Similarly, for $f_i(\vec \sigma(t)) \leq 0$  $\Delta \sigma_i(t) =\mathrm{sgn} \left[-\sigma_i(t) f_i(\vec \sigma(t)) \right]$. Let us define the ramp function $R(x)$ by $R(x) \equiv x$ iff $x \geq 0$ and $R(x) \equiv 0$ iff $x<0$. Using these definitions we can generically map the indicator function $f_i$ from Eq.(\ref{function1})  onto the update equation
\begin{eqnarray}
\sigma_i(t+1) & = & \sigma_i(t)+\Delta \sigma_i(t) \quad, \nonumber \\
\Delta \sigma_i(t) & = &\mathrm{sgn} \left[ \biggl(1-\sigma_i(t) \biggr) R\biggl(f_i(\vec \sigma(t)) \biggr) - \sigma_i(t)  R\biggl(-f_i(\vec \sigma(t))\biggr) \right]
\quad.
\label{DynSys}
\end{eqnarray}

\subsection{Variational principle for deterministic diversity dynamics}

We introduce a distance function to quantify the number of state changes in the system. 
Consider a virtual displacement of $\sigma_i(t)$, $\sigma_i'(t)=\sigma_i(t) + \delta \sigma_i(t)$. 
A quadratic distance function is given by
\begin{equation}
K_i(\sigma_i'(t),\sigma_i(t) ) \equiv \frac{\mu}{2} \left[ \sigma_i'(t) - \sigma_i(t) \right]^2 
\label{kinetic}
\end{equation}
where  $\mu>0$\footnote{Note the similarity to kinetic energy in classical mechanics.}.
Analogously a {\it potential} $V_i$ is defined by
\begin{equation}
V_i(\sigma_i'(t),\vec \sigma(t) ) \equiv \left| \biggl(1-\sigma_i'(t) \biggr) R\biggl(f_i(\vec \sigma(t)) \biggr) - \sigma_i'(t) R\biggl(-f_i(\vec \sigma(t))\biggr)  \right| ,
\label{potential}
\end{equation}
which `counts' the number of possible interactions for the displaced state $\sigma_i'(t)$. Depending on $\sigma_i'(t)$,  Eq.(\ref{potential}) will reduce to $V_i(\sigma_i'(t),\vec \sigma(t)  )=\left| R (\pm f_i(\vec \sigma(t) )) \right|$. A possible intuition behind Eq.(\ref{potential}) is that $f_i$ acts as a `field' on $\sigma_i(t)$ which is `probed' by $\sigma_i'(t)$. 
We occasionally drop the $\sigma$-dependence for a clearer notation. Finally we define the  {\it balance function},  $B_i \equiv K_i+V_i$.

$K_i$ measures the {\it actual} activity in the system -- it counts all state changes. The potential $V_i$ counts the {\it potential} activity in the newly obtained states. $B_i$ therefore contains the full dynamical information of Eq.(\ref{function1}) which can now be expressed through a {\it variational principle}.  

\begin{quote}
{\it Given $\vec \sigma (t)$, the solution $\sigma_i(t+1)$ of Eq.(\ref{DynSys}) is identical to the value of $\sigma_i'(t)$ for which $B_i$ assumes its minimum, i.e. }
\end{quote}
\begin{equation}
\sigma_i(t+1) = \underset{\sigma_i'(t)}{\mathrm{argmin}} \left[ B_i\biggl(\sigma_i'(t),\vec \sigma(t) \biggr) \right]
\quad,
\label{EVP}
\end{equation} 
\begin{quote}
{\it with $\underset{x}{\mathrm{argmin}} \left[ f(x) \right]$ denoting the value of $x$ for which $f(x)$ takes its minimum. }
\end{quote}
This is proved by exhaustive insertion. First, consider the case $f_i(\vec \sigma(t))=0$. From Eq.(\ref{DynSys}) it follows that $\Delta \sigma_i(t)=0$ and $V_i=0$. The only possible term contributing to $B_i$ is $K_i$; $K_i=0$ if $\sigma_i'(t)=\sigma_i(t)$ and $K_i=\frac{\mu}{2}>0$ otherwise. The balance function $B_i$ takes its minimum, $B_i=0$ at $\sigma_i'(t)=\sigma_i(t)=\sigma_i(t+1)$. Similar reasoning can be applied to the cases of non-zero $f_i(\vec \sigma(t))$, see Tab.\ref{proveVP}. This now clarifies the role of the parameter $\mu$. It can be seen as an inertial threshold; the dynamics of Eq.(\ref{DynSys}) only takes place if the `field' $f_i(\vec \sigma(t))$ describing a certain state-change exceeds the barrier set by $\mu$. There is always a choice for $\mu$ such that Eq.(\ref{DynSys}) holds.

\begin{table}
\begin{center}
\caption{We exhaustively insert all possible values for $\sigma_i(t)$ and $\sigma_i'(t)$ in Eq.(\ref{DynSys}) and $B_i$ for non-zero values of $f_i(\vec \sigma(t))$. For convenience we choose  $|f_i(\vec \sigma(t))|=1$ and confine the threshold to $0<\mu<2$. We mark the values of $\sigma_i'(t)$ for which $B_i$ is a minimum by an underscore. 
For these marked values $\sigma_i'(t)$ is always equal to $\sigma_i(t+1)$. }
\label{proveVP}
\begin{tabular}{ccc||ccc||c}
$\sigma_i(t)$ & $f_i(\vec \sigma(t))$& $\sigma_i'(t)$ & $K_i$ & $V_i$ & $B_i$ & $\sigma_i(t+1)$ \\
\hline \hline
0 & -1 & \underline 0 & 0 & 0& $ 0 $ & 0 \\
0 & -1 & 1 & $\frac{\mu}{2}$ & 1 & $1+\frac{\mu}{2}$ & 0 \\
\hline
0 & 1 & 0 & 0 & 1 & 1 & 1 \\
0 & 1 &\underline 1 & $\frac{\mu}{2}$ & 0 & $\frac{\mu}{2}$ & 1\\
\hline
1 & -1 & \underline 0 & $\frac{\mu}{2}$ & 0& $\frac{\mu}{2}$ & 0 \\
1 & -1 & 1 & 0 & 1 & $1$ & 0 \\
\hline
1 & 1 & 0 & $\frac{\mu}{2}$ & 1 & $1+\frac{\mu}{2}$ & 1 \\
1 & 1 &\underline 1 & 0 & 0 & 0 & 1\\
\end{tabular}
\end{center}
\end{table}

\subsection{Stochastic diversity dynamics}

There exists a natural stochastic variant of diversity dynamics. In Eq.(\ref{DynSys}) a state transition $\sigma_i(t)\to \sigma_i(t+1)$ is {\it  determined} by $\Delta \sigma_i(t) \in \{-1,0,1\}$. For the stochastic case we specify {\em transition probabilities} for this evolution.

From the variational principle Eq.(\ref{EVP}) it follows that Eq.(\ref{DynSys}) always minimizes the balance function $B_i$. In the stochastic variant we assume that the lower $B_i$, the higher is the probability to find the system in the respective configuration $\sigma_i(t)$. In analogy to spin systems this probability is a Boltzmann factor 
\begin{equation}
p(\sigma_i(t)) \propto \mathrm e^{-\beta B_i(\vec \sigma(t) )} 
\quad,
\label{boltzmann}
\end{equation}
with $\beta \equiv 1/T$ the inverse temperature. To obtain transition probabilities we demand detailed balance 
\begin{equation}
\frac{p(\sigma_i(t) \to \hat \sigma_i(t))}{p(\hat \sigma_i(t) \to \sigma_i(t))} = \frac{p(\hat \sigma_i(t))}{p(\sigma_i(t))} = \mathrm e^{-\beta (\hat B_i- B_i)}
\quad,
\label{DetBal}
\end{equation}
with $\hat B_i \equiv  B_i(\hat \sigma_i(t),  \sigma(t)_{j \neq i} )$. There are several ways to choose transition probability such that Eq.(\ref{DetBal}) is satisfied, here we use Metropolis transition probabilities $p(\sigma_i(t) \to \hat \sigma_i(t))=1$ if $\hat B_i- B_i<0$ and $p(\sigma_i(t) \to \hat \sigma_i(t))=\exp[-\beta (\hat B_i- B_i)]$ otherwise. The stochastic diversity dynamics is fully specified by setting\footnote{One can also define dynamics `backwards' by the transition probability $p(\hat \sigma_i(t) \to \sigma_i(t))$ which can be interpreted as inferring $\vec \sigma(t)$ from the knowledge of $\vec \sigma(t+1)$.}
$\sigma_i(t+1) = \hat \sigma_i(t)$.

Whereas in the deterministic case the balance function $B_i$ is minimized, the stochastic diversity dynamics shows `disordering effects' due to non-zero temperature $T$ as given in Eq.(\ref{boltzmann}). We quantify this with Boltzmann-Gibbs entropy.

\subsection{Mean-field approximation}

Denote the expectation value of $\sigma_i(t)$ by $q_i(t)=\langle \sigma_i(t) \rangle$ and assume that the probability distribution factorizes, i.e. $p(\vec \sigma(t))=\prod_i p_i (\sigma_i(t))$ with $p_i(\sigma_i(t))=(1-q_i(t)) \delta_{\sigma_i(t),0}+q_i(t) \delta_{\sigma_i(t),1}$. In this mean-field approximation the Boltzmann-Gibbs entropy $s$ for species $i$ is 
\begin{eqnarray}
s(\sigma_i(t)) & = & - \langle \ln p_i(\sigma_i(t)) \rangle \equiv s(q_i(t)) \quad, \\
s(q_i(t)) & = &-\biggl(1-q_i(t) \biggr) \ln (1-q_i(t)) -q_i(t) \ln q_i(t) \nonumber \quad.
\label{entropy}
\end{eqnarray}
The `free energy' functional $\phi(\sigma_i(t))$ for the system turns for this approximation into
\begin{equation}
\phi(q_i(t)) = \langle B_i \rangle_{p(\vec \sigma(t))} -\frac{s(q_i(t) )}{\beta}
\quad.
\label{phi}
\end{equation}
The asymptotic state of species $i$, $q_i(t \to \infty) \equiv q_i$, is identified by a minimum in free energy. The necessary condition for this, $\partial \phi(q_i) / \partial q_i =0$, is $\frac{\partial \langle B_i \rangle}{\partial q_i} + \frac{1}{\beta} \ln \left(\frac{q_i}{1-q_i} \right) = 0 $, and
\begin{equation}
q_i = \frac{1}{2} \left\lbrace \tanh \left[- \frac{\beta}{2} \frac{\partial \langle B_i \rangle}{\partial q_i} \right] +1 \right\rbrace 
\quad.
\label{MFsol}
\end{equation}
The self-consistent solution of Eq.(\ref{MFsol}) yields the asymptotic configuration.

\section{General formulation of evolutionary interactions}
\label{sec: MotivMod}
Traditionally in the master equations framework\footnote{as is typical for traditional evolutionary biology.} interactions are classified by transfer rates for abundances of species. 
The transfer rates measure how the change in abundance of a given species $i$ is related to the abundance of other species $j_1, j_2, \dots$.
Depending on how $i$ and $j_1, j_2, \dots$ are chosen, one obtains different systems of differential equations which can be related to a specific form of evolutionary interactions.
If species $i$ with abundance $x_i$ replicates with rate $f_i$,  the interaction is of type {\it replication} and is represented as $x_i \overset{f_i}{\to} 2 x_i$, (replicator equation \cite{Crow70}). 
{\it Competition} is a mechanism where the replication rate of species $i$ also depends on other species $j$ through a transfer rate $p_{ij}$, $x_i+x_j \overset{p_{ij}}{\to} x_i$. This type of interactions is used in the game dynamical equation \cite{Schuster83}, which is a special case of the frequency dependent replicator equation \cite{Hofbauer98, Taylor78}. 
The mechanism {\it mutation} assigns a mutation or transfer rate $q_{ij}$ between two species according to $x_i \overset{q_{ij}}{\to} x_j$, together with replication and competition we obtain the replicator-mutator equation \cite{Hadeler81}, of which the quasispecies equation \cite{Eigen89} is a special case. Replication can take place without replicators, species are then produced by {\it recombination} processes. In the case of three species $i$, $j$ and $k$ with a recombination rate $\alpha_{ijk}$, this mechanism is $x_j+x_k \overset{\alpha_{ijk}}{\to} x_i$. The corresponding dynamical system is called {\em catalytic network} see e.g. \cite{Stadler93}. It is formally possible to express replication, mutation and competition as special cases of the recombination mechanism \cite{Fontana94, Page02}. In this sense recombination mechanisms provide a unifying description of the other evolutionary interactions above -- an observation we use  as a starting point for our model.

In the general form of a recombination process an arbitrary number of species $j_1, j_2, \dots, j_n$ influences a given species $i$. We distinguish two types of interactions of this form, (i) {\em constructive} interactions or productions where species $i$ benefits from species $j_1, j_2, \dots, j_n$ and (ii) {\em destructive} interactions or destructions where the $j$'s are causing harm to $i$. In the master equation framework constructive interactions correspond to positive transfer rates, destructions to negative ones. We denote the set of species $j_1, j_2, \dots, j_n = \mathbf{j}$. If the set of all species is $\mathcal N$, {\bf j} is an element of the set of all subsets of $\mathcal N$, i.e.  {\bf j} is an element of the power set of $\mathcal N$, $\mathcal P ( \mathcal N )$. A recombination always maps an element from $\mathcal P ( \mathcal N )$ to an element from $\mathcal N$ via a transfer rate $\alpha_{i, \mathbf j}$, i.e. by  a map $\alpha : \mathcal P (\mathcal N) \to \mathcal N$. From now on italic indices refer to elements of $\mathcal N$, e.g. $i \in \mathcal N$, while bold-face indices refer to elements of the power set, $\mathbf j \in \mathcal P (\mathcal N)$. 
Transfer rates are represented by their sign.
For convenience define binary state variables for sets of nodes, let $\sigma_{\mathbf j}(t) = \prod_{i \in \mathbf j}\sigma_i(t)$. The most general form of evolutionary interactions can then be written as 
\begin{equation}
\sigma_{\mathbf j} \overset{\alpha_{i, \mathbf j}}{\to} \sigma_i
\quad.
\label{EvoGen}
\end{equation} 
We summarize in Tab.\ref{axioms} how the evolutionary interaction mechanisms of replication, competition, mutation and recombination are contained in Eq.(\ref{EvoGen}) for special choices of {\bf j}. If there is a constructive interaction between $\mathbf j$ and $i$, i.e. {\bf j} is a {\it constructive set}, we capture it in the {\it production rule table} $\alpha^+ : \mathcal P (\mathcal N) \to \mathcal N$ with  $\alpha^+_{i, \mathbf j}=1$, otherwise $\alpha^+_{i, \mathbf j}=0$. Similarly, if the interaction in Eq.(\ref{EvoGen}) is destructive, i.e. {\bf j} is a {\it destructive set}, we record this in the {\it destruction rule table} $\alpha^- : \mathcal P (\mathcal N) \to \mathcal N$ with  $\alpha^-_{i, \mathbf j}=1$, otherwise $\alpha^-_{i, \mathbf j}=0$.
At some points in this work we will assume that the rule tables $\alpha^{\pm}$ are random tensors. In this case they are given by two parameters $n^{\pm}$ and $r^{\pm}$. $n^+$ is the cardinality of constructive sets, $\vert \mathbf j \vert = n^+$ in Eq.(\ref{EvoGen}) and for random $\alpha^+$ each species $i$ has on average the same number $r^+$ of constructive sets, $\langle \sum_{\mathbf j} \alpha^+_{i, \mathbf j} \rangle_i=r^+$. Similarly $\alpha^-$ is given by $n^-$ and $r^-$.

\begin{table}
\caption{Summary of the traditional evolutionary interaction mechanisms: replication, competition, mutation and recombination. We indicate how the constructive/destructive set {\bf j} has to be specified in Eq.(\ref{EvoGen}) in order to recover the various mechanisms  in our model. }
\begin{center}
\begin{tabular}{lcrclcl}
mechanism & & &   & & & power set notation \\
\hline
replication &\hspace{0.5cm} & $x_i$ & $\overset{f_i}{\to}$ &  $2x_i$ & \hspace{0.5cm} &{\bf j}$=\{i\}$ \\
competition &  &$x_i+x_j$ &  $\overset{p_{ij}}{\to}$ & $x_i$ &  &{\bf j}$=\{i,j\}$ \\
mutation & & $x_j$ & $\overset{q_{ij}}{\to}$ & $x_i$& & {\bf j}$=\{j\}$ \\
recombination & & $x_j+x_k$ & $\overset{\alpha_{ijk}}{\to}$& $x_i$ & &{\bf j}$=\{j,k\}$ \\
\end{tabular}
\end{center}
\label{axioms}
\end{table}

\begin{figure}[t]
\begin{center}
\includegraphics[height=4.5cm]{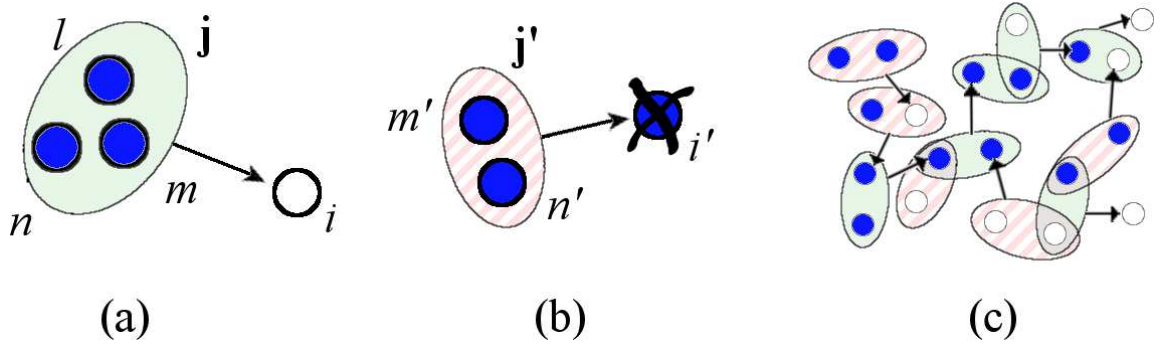}
\end{center}
\caption{A graphical representation of constructive, destructive and combined interactions. (a) The constructive set {\bf j} (green area) contains three species (circles) $\mathbf j = \{l,m,n\}$. They produce species $i$, i.e. $\alpha_{i, \mathbf j}^+=1$. The constructive set is active (indicated by the blue color of the circles) we have $\sigma_{\mathbf j}(t)=1$ and $\sigma_i(t)=0$, by Eq.(\ref{ConDyn}) $\sigma_i(t+1)=1$. (b) The destructive set {\bf j'} (red striped area) of cardinality two is active since each of its contained species $m', n'$ is active and interacting with species $i'$ through $\alpha_{i', \mathbf {j'}}^-=1$. Following Eq.(\ref{DesDyn}), $\sigma_{i'}(t)=1$ will be deactivated, $\sigma_{i'}(t+1)=0$. (c) A pictorial description of a network with both constructive and destructive interactions at a point in time.}
\label{balls}
\end{figure}

\section{Constructive interactions}
\label{sec: Constr}

\subsection{Constructive dynamical system}

We first consider a system with constructive interactions only. We read Eq.(\ref{EvoGen}) as `from $\sigma_{\mathbf j}(t)=1$ follows that $\sigma_i(t+1)=1$'. In a chemical setting the chemical compounds contained in {\bf j} react to give compound $i$, in an economic setting the goods {\bf j} can be assembled to produce good $i$, see Fig.\ref{balls}(a). The constructive dynamical system characterized by Eq.(\ref{EvoGen}) is given by
\begin{equation}
\Delta \sigma_i(t) = \mathrm{sgn} \left( \biggl( 1-\sigma_i(t) \biggr) \sum_{\mathbf j \in \mathcal P ( \mathcal N)} \alpha_{i, \mathbf j}^+ \sigma_{\mathbf j}(t) \right)
\quad,
\label{ConDyn}
\end{equation}
i.e. $f_i(\vec \sigma(t))$ from Eq.(\ref{function1}) becomes $f_i^+(\vec \sigma(t))=\sum_{\mathbf j} \alpha_{i, \mathbf j}^+ \sigma_{\mathbf j}(t)$.

\subsection{Deterministic constructive diversity dynamics}

In the limiting case of $T=0$ i.e. $\beta \to \infty$ the system deterministically obeys the dynamics of Eq.(\ref{ConDyn}). The behavior of $D(t \to \infty)$ is well understood; this case is identical to the model studied in \cite{htk1, htk2} for random interaction topologies $\alpha^+$ given by $n^+,r^+$ here.
 $D(\infty)$ was computed as a function of $n^+, r^+$ {\it and} $D(0)$. It was shown that  this system has a phase transitions formally equivalent to the phase transition of a van der Waals Gas. There exists a critical diversity of initial species $D_c(0)$ above which the system is driven toward an almost  fully populated state; below this threshold the dynamics freezes. All these findings are identical what we find here in the $T=0$ case.

\subsection{Stochastic constructive diversity dynamics}

We next turn to non-zero temperature $T$. The crucial feature distinguishing deterministic and stochastic diversity dynamics is the {\it dependence on the initial conditions}. In the presence of stochastic perturbations the final diversity is not a function of the initial diversity $D(0)$. We employ a mean-field approach by assuming that the expectation value of a product equals the product of expectation values,
$\langle g_1 (\sigma) g_2 (\sigma) \rangle =  \langle g_1 (\sigma) \rangle \langle g_2 (\sigma) \rangle$.
The expectation value of the constructive potential $\langle V^+_i \rangle_{p(\vec \sigma(t))}$ of species $i$ is
\begin{equation}
\langle V^+_i \rangle_{p(\vec \sigma(t))} = \biggl(1-q_i(t)\biggr) \sum_{\mathbf j \in \mathcal P ( \mathcal N)} \alpha_{i, \mathbf j}^+ \prod_{j \in \mathbf j} q_j(t)
\quad,
\label{ConPot}
\end{equation}
quantifying what {\it could} be produced given the actual configuration of the system\footnote{Note that the structure of $V^+_i$ has a strong similarity to the potential of the paradigmatic Ising model. Our model diverges in the following ways: (i) interactions are defined not between nodes but between constructive sets and nodes and (ii) interactions are not symmetric, the action of {\bf j} on $i$ does not equal the action of $i$ on {\bf j}.}.

\begin{figure}[t]
\begin{center}
\includegraphics[height=3.8cm]{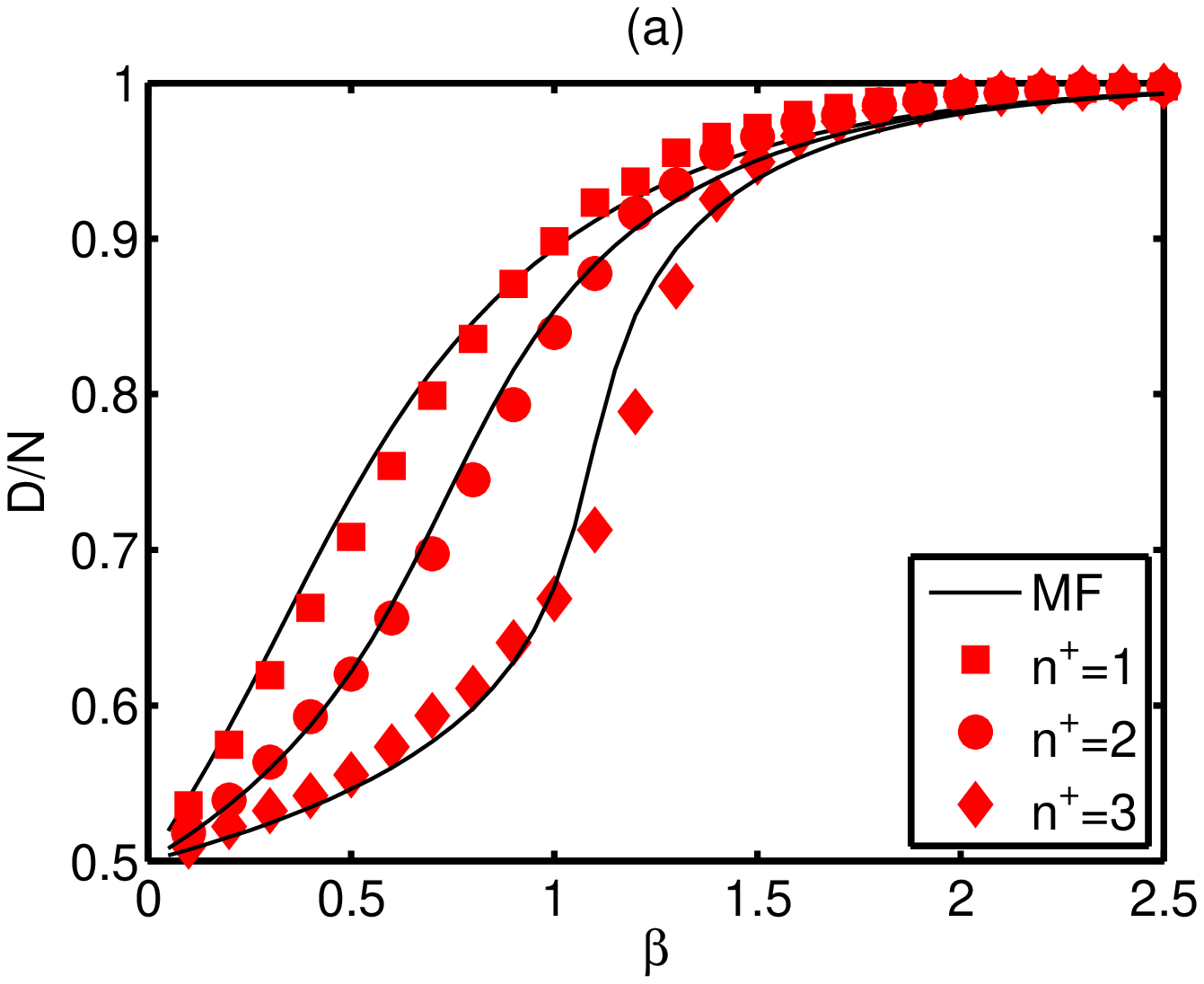}
\includegraphics[height=3.8cm]{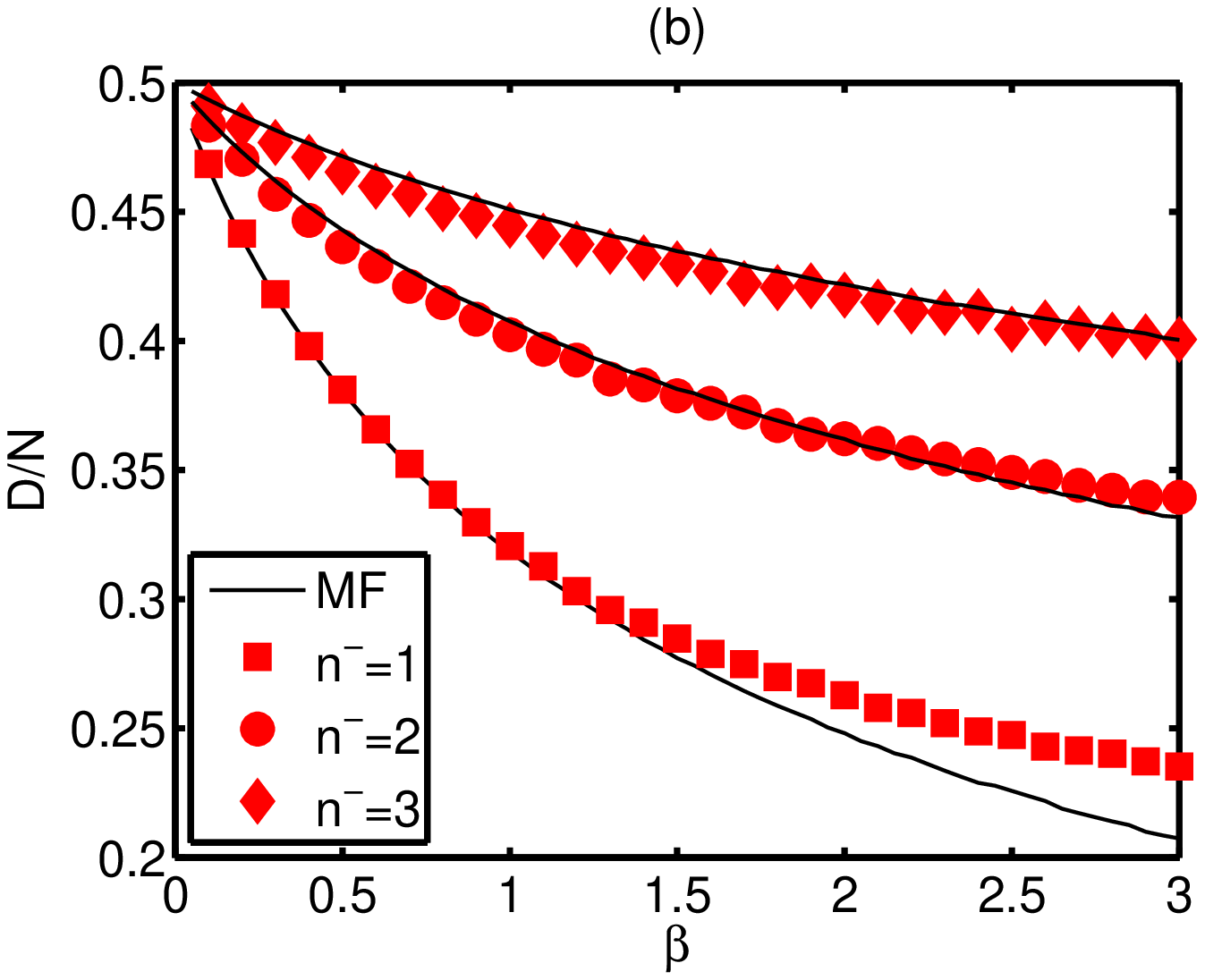}
\includegraphics[height=3.8cm]{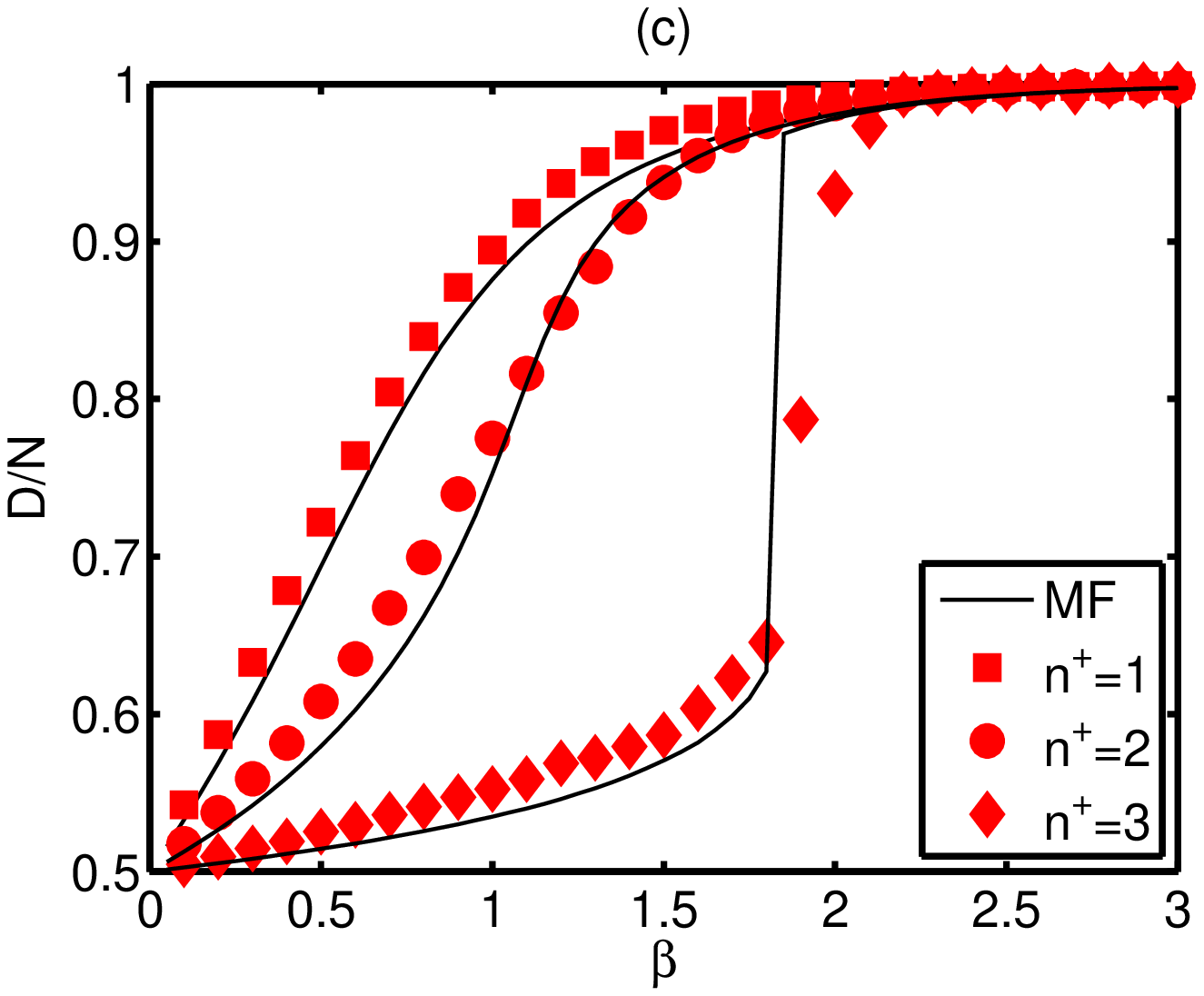}
\end{center}
\caption{Diversity as a function of inverse temperature $\beta$ for various dynamical systems obtained from a mean-field approach (MF, lines) and Metropolis simulations (symbols). (a) Constructive dynamics with $r^+=2$. (b) Destructive dynamics with $r^-=2$. (c) Combined dynamics with $r^+=3$, $r^-=1$ and $n^-=2$.}
\label{magncurv}
\end{figure}

As mentioned above,  $\mu$ plays the role of a threshold. For $0<\mu<2$ a species gets activated by one constructive set, for $2<\mu<4$ at least two constructive sets are needed and etc. From now on we fix the threshold $\mu=1$. The contribution to free energy is $K^+_i(t)=\frac{1}{2} (\Delta \sigma_i(t))^2$. We can estimate the expectation value $\langle K_i(t) \rangle_{p(\vec \sigma(t))}$ by making use of the dynamical relation Eq.(\ref{ConDyn}), and get for the  mean-field assumption 
\begin{equation}
\langle K^+_i \rangle_{p(\vec \sigma(t))} =\frac{1}{2} \left( \biggl( 1-q_i(t) \biggr) \sum_{\mathbf j \in \mathcal P ( \mathcal N)} \alpha_{i, \mathbf j}^+ \prod_{j \in \mathbf j} q_j(t) \right)^2
\quad.
\label{ConKin}
\end{equation}
Using this in Eq.(\ref{MFsol}) gives us the mean-field solution for arbitrary interaction topologies $\alpha^+$. To compute it explicitly we assume  random interaction topologies. 
The aim is to derive an expression for $\frac{\partial \langle B_i \rangle}{\partial q_i}$ in the limit $t \to \infty$. Note that $\frac{\partial \langle V_i \rangle}{\partial q_i}=-\sum_{\mathbf j \in \mathcal P ( \mathcal N)} \alpha_{i, \mathbf j}^+ \prod_{j \in \mathbf j} q_j$. Due to the randomness in $\alpha^+$ the same average `field' is exerted on each species. With $q \equiv \langle q_i \rangle_i$ we get $\frac{\partial \langle V_i \rangle}{\partial q_i}=-r^+ q^{n^+}$.
We apply the same reasoning to the distance-contribution $\langle K^+_i \rangle_{p(\sigma(t))}$. By  first carrying out the derivation and then putting in the  assumptions about $\alpha^+$, we get
\begin{equation}
\frac{\partial \langle B_i \rangle}{\partial q_i} = -r^+ q^{n^+} - (1-q) (r^+ q^{n^+})^2
\quad,
\label{ConB}
\end{equation}
and the self-consistent solution for the asymptotic abundance $q$,
\begin{equation}
q = \frac{1}{2} \left\lbrace \tanh \left[\frac{\beta}{2}\left( r^+ q^{n^+} + (1-q) (r^+ q^{n^+})^2 \right) \right] +1 \right\rbrace 
\quad,
\label{qSolCon}
\end{equation}
from which the diversity follows as $D(t \to \infty)=Nq$. We compare predictions of Eq.(\ref{qSolCon}) with simulations results from a Metropolis algorithm. The latter was implemented in the following way:  We constructed a random $\alpha^+$ and initialized the system with a random initial condition $\vec \sigma(0)$. After initialization the algorithm applies the  following procedure to each species once within one timestep (random sequential update):
\begin{itemize}
\item Pick a species $i$ randomly.
\item Calculate $B_i=K_i+V_i$, according to Eqs.(\ref{kinetic}) and (\ref{potential}) with $\sigma_i'(t)=\sigma_i(t)$.
\item Calculate $B_i=K_i+V_i$ with $\sigma_i'(t)=1-\sigma_i(t)$.
\item Calculate $\Delta B=B_i \biggl(\sigma_i'(t)=1-\sigma_i(t)\biggr)-B_i \biggl( \sigma_i'(t)=\sigma_i(t) \biggr)$ 
\item If $\Delta B <0$ set $\sigma_i(t+1)=1-\sigma_i(t)$.
\item If $\Delta B >0$ set $\sigma_i(t+1)=1-\sigma_i(t)$ with probability $\mathrm e^{-\beta \Delta B}$.
\end{itemize} 
We executed the algorithm  for one particular realization of $\alpha^+$  for  $10^3$ timesteps
 and averaged over this time-span after discarding transient behavior (typically about 50 iterations). We performed simulations for system sizes  of $N=10^2-10^4$ without noticing size effects on the results. However, the time-to-converge depends on $N$. We show the degree of agreement of simulations and Eq.(\ref{qSolCon}) in Fig.\ref{magncurv}(a).

\section{Destructive dynamics}
\label{sec: Destr}
\subsection{Destructive dynamical systems}

Assume now that only destructive interactions take place, e.g. two chemicals catalyzing the consumption of another chemical species, or biological species gaining (in symbiosis) an evolutionary advantage over another species. Eq.(\ref{EvoGen}) is now read as `from $\sigma_{\mathbf j}(t)=1$ follows $\sigma_i(t+1)=0$', see Fig.\ref{balls}(b). To formulate this as a dynamical system as in Eq.(\ref{DynSys}) set $f_i(\vec \sigma(t)) \rightarrow f^-_i(\vec \sigma(t)) =-\sum_{\mathbf j } \alpha_{i, \mathbf j}^- \sigma_{\mathbf j} (t) $ to get 
\begin{equation}
\Delta \sigma_i(t) = \mathrm{sgn} \left( -\sigma_i(t) \sum_{\mathbf j \in \mathcal P (\mathcal N)} \alpha_{i, \mathbf j}^- \sigma_{\mathbf j} (t) \right)
\quad.
\label{DesDyn}
\end{equation}
We  discuss the deterministic ($T=0$) and stochastic ($T>0$) scenario.

\subsection{Deterministic destructive diversity dynamics}

In the deterministic case the asymptotic diversity $D(t \to \infty)$ is a function of the initial diversity. Let us discuss the case of a completely random destructive rule table $\alpha^-$. By denoting $q(t)=D(t)/N$ we can derive an update equation for $q(t)$ following the same reasoning as in \cite{htk1}, $q(t+1)=q(t)-\Delta^- q(t)$ with $\Delta^- q(t) = r^- q(t) \biggl( q^{n^-}(t)-q^{n^-}(t-1) \biggr)$. In the limit of sparse rule densities $r^-$ this leads to  $q(t \to \infty)=q(0)-n^- r^- q^{n^- +1}$.
In contrast to constructive dynamics, destructive dynamics do not exhibit a phase transition. With more species being destroyed the number of deactivated destructive sets increases even faster, thus the process comes to a halt without reaching a strongly  unpopulated state.
Note that other kinds of destructive dynamics may exhibit phase transitions.
For example, if one requires for each species to be abundant at least one productive set to be abundant, that is species become extinct once they are not actively produced, the removal of a small number of species may trigger a cascade of extinction events which erases the entire population \cite{htk2}.

\subsubsection{Stochastic destructive diversity dynamics}

For a stochastic variant of the destructive dynamical system of Eq.(\ref{DesDyn}) we repeat the analysis of the constructive case. We start with the corresponding destructive potential and distance terms,
\begin{eqnarray}
\langle V^-_i \rangle_{p(\vec \sigma(t))} & = & q_i(t) \sum_{\mathbf j \in \mathcal P ( \mathcal N)} \alpha_{i, \mathbf j}^- \prod_{j \in \mathbf j} q_j(t) \quad, \nonumber \\
\langle K^-_i \rangle_{p(\vec \sigma(t))} & = & \frac{1}{2} \left( q_i(t) \sum_{\mathbf j \in \mathcal P ( \mathcal N)} \alpha_{i, \mathbf j}^- \prod_{j \in \mathbf j} q_j(t) \right)^2
\quad.
\label{DesPotKin}
\end{eqnarray}
We proceed with the derivation of the destructive balance function $B_i$ and get 
\begin{equation}
\frac{\partial \langle B_i \rangle}{\partial q_i} = r^- q^{n^-} + q (r^- q^{n^-})^2
\quad,
\label{DesB}
\end{equation}
and the self-consistent solution for the asymptotic abundance $q$,
\begin{equation}
q = \frac{1}{2} \left\lbrace \tanh \left[- \frac{\beta}{2}\left( r^- q^{n^-} + q (r^- q^{n^-})^2 \right) \right] +1 \right\rbrace 
\quad.
\label{qSolDes}
\end{equation}
We compare this prediction to results of a Metropolis simulation in Fig.\ref{magncurv}(b). 
As is seen in the $n^-=1$ case, the deviation between Eq.(\ref{qSolDes}) and simulations increases with $\beta$. For higher $n^-$ and $r^-$ the same extent of deviation occurs for a higher value of $\beta$. The mean-field approximation starts to significantly differ from simulations once entropic effects become negligible and the system's evolution approaches the deterministic scenario, that is  $\mathrm e^{-\beta \Delta B_i} \lesssim 1/N$ $\forall i$ (on average less than one random state flip per iteration). To approximate $V^-_i(t)=\sigma_i(t) \sum_{\mathbf j} \alpha_{i, \mathbf j}^- \prod_{j \in \mathbf j} \sigma_j(t)$ at any time $t$ we have to consider the species which have {\it not} been deactivated at $t-1$ -- the system possesses memory. This is not captured in the mean-field approximation $\langle V^-_i \rangle= r^- q^{n^-}$ where we assume the populated species to be randomly distributed over $N$ possible species at each time $t$. In the destructive case the mean-field approach thus works best whenever the random fluctuations are large enough to `smear out' this memory effect, otherwise the system is better approximated by the deterministic description.

\section{Combined dynamics}
\label{sec: CombDyn}

\subsection{Combined dynamical systems}
We now study the interplay of both constructive and destructive dynamics \cite{thurner09, tkh09}; the situation is sketched in Fig.\ref{balls}(c). Destructive interactions represent an implicit  selection mechanism\cite{tkh09}. Each species may be targeted (influenced) by constructive {\it and} destructive interactions. Assume that each interaction has equal influence. If the constructive forces outweigh the destructive ones the species prefers to be active and {\it vice versa}. For some systems other choices of weighting could be more appropriate (e.g. assuming that {\it one} destructive interaction outweighs any number of constructive ones -- `it is easier to destroy than to build'). It is straight-forward to incorporate alternative weighting schemes in the present framework.

To combine constructive and destructive interactions we  add their indicator functions,
\begin{equation}
f_i(\vec \sigma(t)) = f_i^+(\vec \sigma(t))+f_i^-(\vec \sigma(t))= \sum_{\mathbf j} \alpha_{i, \mathbf j}^+ \sigma_{\mathbf j}(t)-\sum_{\mathbf j} \alpha_{i, \mathbf j}^- \sigma_{\mathbf j}(t)
\quad,
\label{delta}
\end{equation}
and get for the dynamical equation
\begin{equation}
\Delta \sigma_i(t) = \mathrm{sgn} \left[ \biggr(1-\sigma_i(t) \biggr) R(f_i(\vec \sigma(t)) )-\sigma_i(t) R(-f_i(\vec \sigma(t)) ) \right]
\quad.
\label{CombDyn}
\end{equation}
The purely destructive or constructive dynamical systems are recovered by setting $\alpha^{\pm}=0$.

\subsection{Deterministic combined diversity dynamics}
To obtain an estimate for the asymptotic diversity, 
we again use an update equation and combine the finding for the constructive and destructive cases. 
If we denote the average in-(de)crements in the constructive (destructive) scenario by  $\Delta q^+(t)$ ( $\Delta q^-(t)$), we  study the update equation $q(t+1)=q(t)+\Delta q^+(t)-\Delta q^-(t)$. This equation is solved by using the same Ansatz as in \cite{htk1}, yielding $q(t \to \infty)=q(0)-n^- r^- q^{n^- +1}+n^+ r^+ (1-q) q.^{n^+}$.

\subsection{Stochastic combined diversity dynamics}

Let us calculate $\langle B_i \rangle$ for the stochastic scenario. The expectation value of the distance contribution, $\langle K_i \rangle_{p(\vec \sigma(t))}$, is more involved now. Constructive (destructive) dynamics take place under the condition that $f_i(\vec \sigma(t)) \geq 0 (\leq 0)$. Start with an expression for the probability that $f_i(\vec \sigma(t))$ is positive (negative), $p^{\pm}$. 
Consider random interaction topologies specified by $r^{\pm}$ and $n^{\pm}$.
Define $p(k,r^+)$ as  the probability that there are exactly $k$ active constructive interactions, that is $p(k,r^+) \equiv  {r^+ \choose k} q^{n^+k} (1-q^{n^+})^{r^+-k}$. 
Analogously, $q(l,r^-)$ is the probability that exactly $l$ out of $r^-$ destructive interactions are active. Then
\begin{eqnarray}
p^+ & = & \sum_{k=1}^{r^+} p(k,r^+) \sum_{l=0}^{\min (k-1,r^-)} q(l,r^-) \quad, \nonumber \\
p^- & = & \sum_{l=1}^{r^-} q(l,r^-) \sum_{k=0}^{\min (l-1,r^+)} p(k,r^+)  \quad.
\label{probs}
\end{eqnarray}
The average  distance follows as
\begin{equation}
\langle K_i \rangle_{p(\sigma)}= \frac{1}{2} \biggl( (1-q_i) p^+ +q_i p^- \biggr)^2
\quad,
\label{CombKin}
\end{equation} 
and, abbreviating $f_i(\vec \sigma(t)) \equiv f_i$, the potential is
\begin{equation}
 \langle V_i \rangle_{p(\sigma)}=\left| \left( 1-q_i \right) R( f_i) -q_i R(-f_i) \right|
\quad.
\label{CombPot}
\end{equation}
Taking the derivative with respect to $q_i$ the mean-field result is
\begin{equation}
\frac{\partial \langle B_i \rangle}{\partial q_i} = -r^+q^{n^+} +r^- q^{n^-} - \left[ (1-q) p^+ + q p^- \right] (p^+ - p^-)
\quad,
\label{CombB}
\end{equation}
with the self-consistent solution for the asymptotic abundance $q$
\begin{equation}
q = \frac{1}{2} \left\lbrace \tanh \left[\frac{\beta}{2}\left(r^+q^{n^+} -r^- q^{n^-} + \left[(1-q) p^+ + q p^-\right](p^+ - p^-) \right) \right] +1 \right\rbrace 
.
\label{qSolComb}
\end{equation}
Again we compare the mean-field prediction to results of a Metropolis simulation of the full model in Fig.\ref{magncurv}(c).

\begin{figure}[t]
\begin{center}
\includegraphics[height=3.8cm]{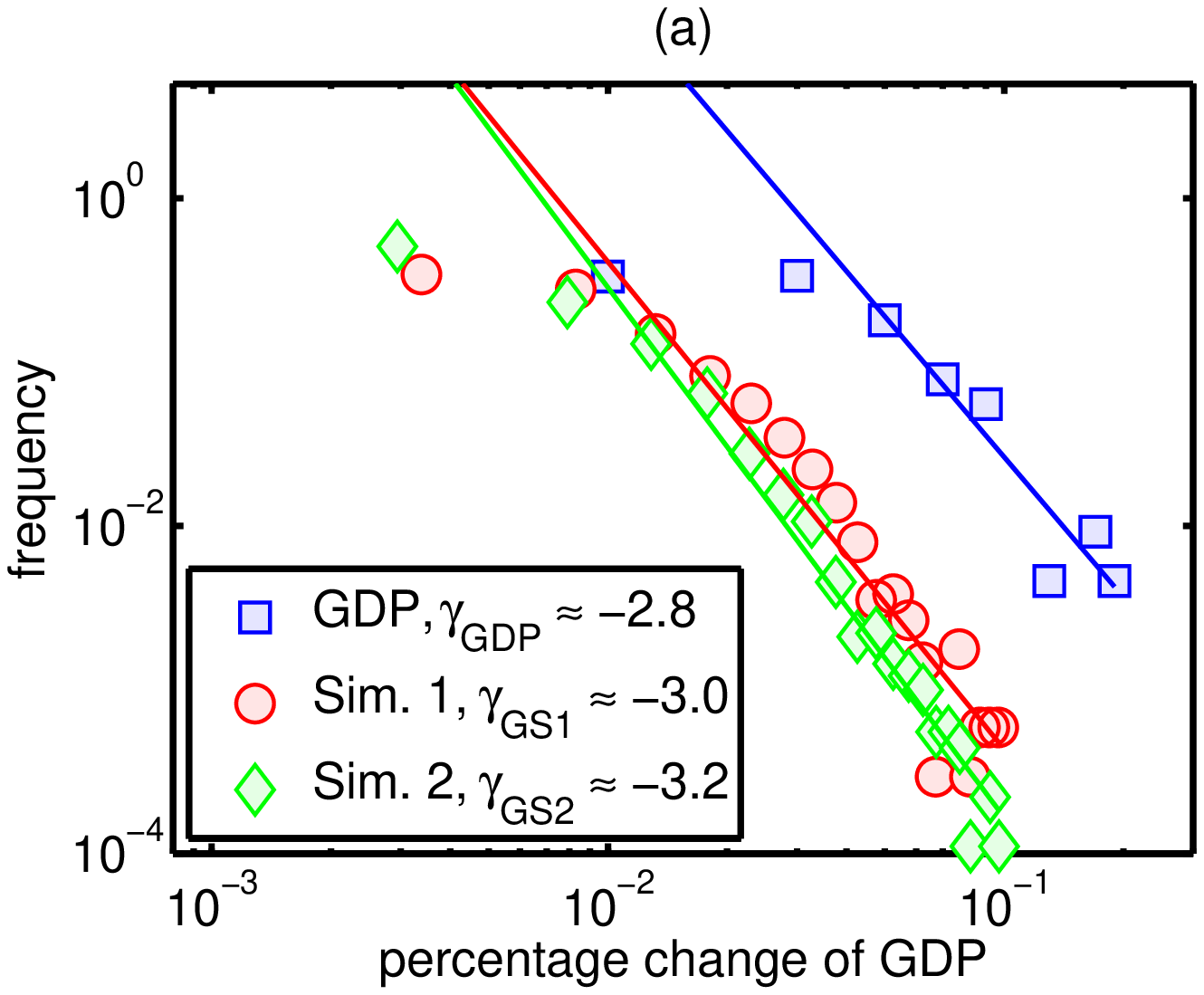}
\includegraphics[height=3.8cm]{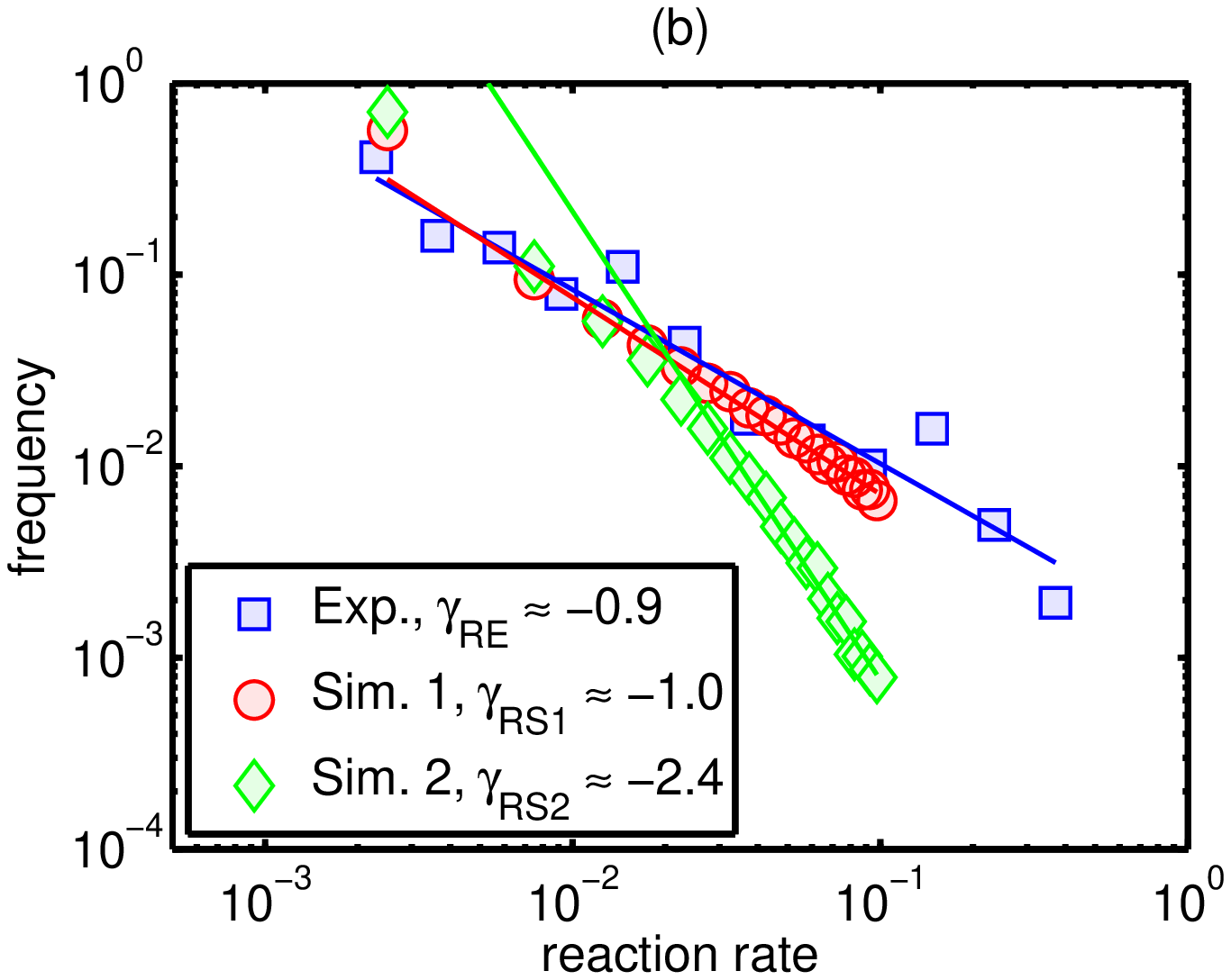}
\includegraphics[height=3.8cm]{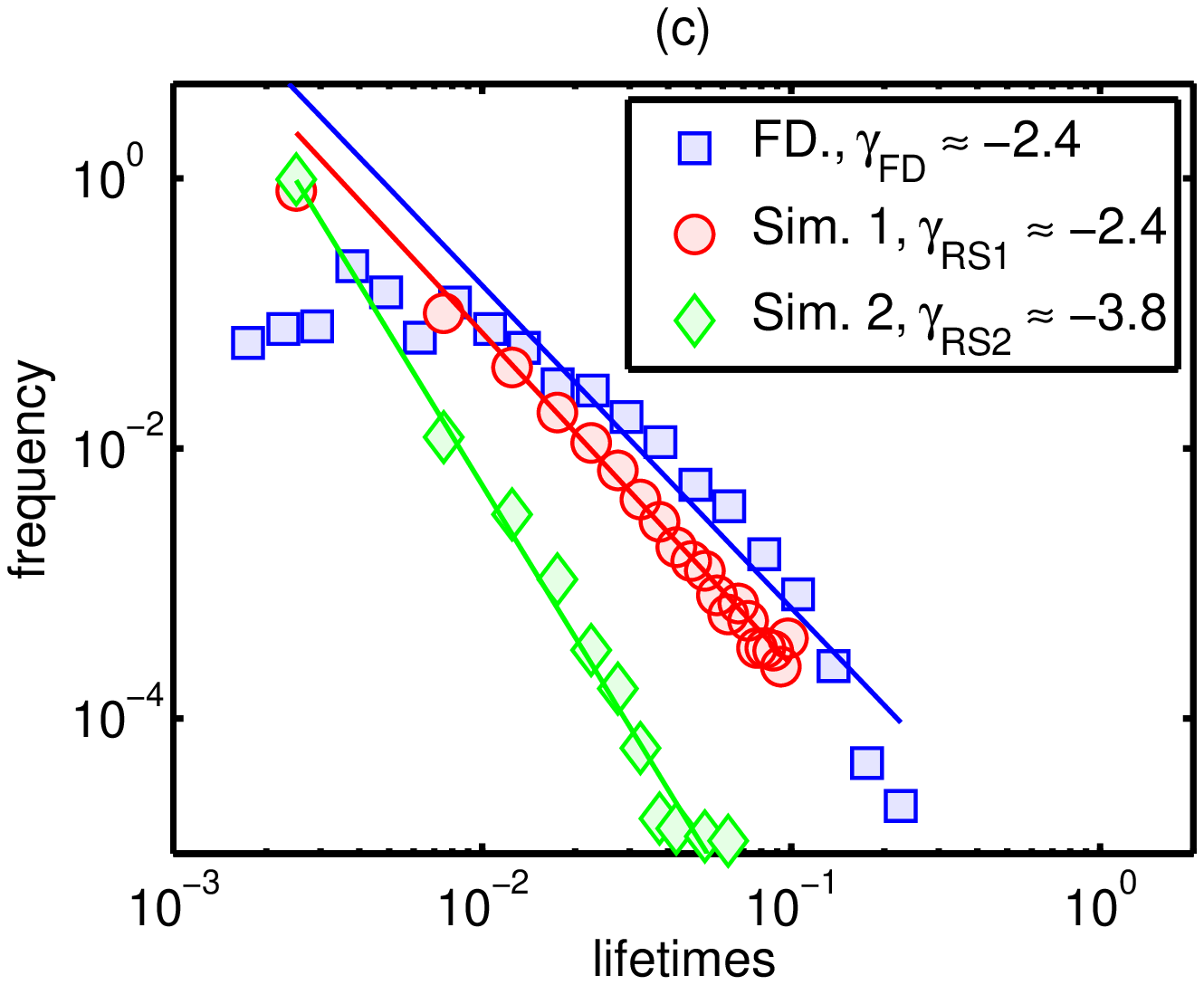}
\end{center}
\caption{We compare the distribution of systemic observables of evolutionary systems with those of the combined stochastic  model for two different parameter settings: {\em Simulation 1} with $\beta=15$, $r^{\pm}=5$, $n^{\pm}=2$ and {\em Simulation 2} with $\beta=15$, $r^+=8$, $r^-=12$, $n^{\pm}=2$. Each distribution has been normalized (sum over all data points equals one). (a) The percent change of GDP of the UK since 1950 is compared to the model. (b) The reaction rate distribution in the model and in the metabolic network of {\it E. coli} is shown. (c) Species lifetime distributions as found in fossil data are well reproduced with the model.}
\label{data}
\end{figure}

\section{Discussion on Empirical Relevance}
\label{sec: EmpRel}

\subsection{Economical setting}
We interpret the model in different evolutionary contexts and compare its behavior to measured data. In an economic setting one can identify the number of active interactions as a measure for the productive output of an economy -- for example the GDP \cite{tkh09}. An interaction is defined to be active iff $\sigma_i(t)=\sigma_{\mathbf j}(t)=\alpha_{i,\mathbf j}^{\pm}=1$. We show  in Fig.\ref{data}(a) a comparison between the actual distribution of percent increments of the GDP of the UK  and  the number of active productions from the combined stochastic model for two different parameter settings. In one setting  $\beta=15$, $r^{\pm}=5$, $n^{\pm}=2$ is used, the other has a denser interaction topology, $\beta=15$, $r^+=8$, $r^-=12$, $n^{\pm}=2$.  Both model and real-world  GDP  timeseries produce fat-tailed distributions, with power exponents in the range between -2 and -4. These features are also found in GDP timeseries of other countries and for a wide range of model parameters, see e.g. \cite{tkh09}.

\subsection{Chemical setting}
Another possible interpretation of the combined stochastic system is a chemical reaction network. In this case chemical species $\mathbf j = \{j_1, j_2, \dots \}$ are producing or degrading chemical $i$. There are $N(r^+ + r^-)$ reactions. A reaction rate is defined as the frequency with which a certain reaction is active and a reaction is active if $\alpha^{\pm}_{i,\mathbf j}=1$, $\sigma_i(t)=1$ and $\sigma_{\mathbf j}(t)=1$. This is compared to reaction rates in the metabolic network of {\it E. coli} \cite{emmerling02} in Fig.\ref{data}(b). Distributions of reaction rates in both cases, model and living organism, are fat-tailed.  Least-squares fits to model power-laws yield exponents in the range of  $-1$ to $-3$, depending on parameters.  This compares well  to the value of $-1$ found for {\it E. coli}. 

\subsection{Biological evolution setting}
Translated into a macro-ecological setting, one can compare the distribution of lifetimes of species in the combined stochastic model (number of iterations a given species is unintermitted abundant) with the distribution of species lifetimes in fossil data \cite{Sepkoski92} in Fig.\ref{data}(c). Again one finds power-laws  in the model with exponents between $-2$ and $-4$, which  matches well with the paleontologic  data, which suggest slopes between $-2$ and $-3$. Note that there is a strong  dependence on the values used for the fit. We work with an intermediate choice in Fig.\ref{data}(c). 

\section{Discussion}
\label{sec: Discuss}
We propose a general framework to systematically study a large  class of dynamical evolutionary systems defined on an arbitrary large number of species. 
The trajectory of existence of each species is governed by a function incorporating information of the surroundings -- the existence of other species.  
We show how to express the resulting system dynamics via a variational principle. We discuss deterministic and stochastic variants. For the latter we derive a closed expression for the asymptotic diversity of evolutionary systems within a mean-field approximation. We discuss the quality of this approximation with respect to Metropolis simulations of the full model. 
Although the model explicitly introduces strong correlations between species' abundances, the mean-field approximation for asymptotic diversities match the simulation data surprisingly well. 
The model can be seen as a generalization of several previous models, which are contained as special cases.
The deterministic constructive case is identical to the random catalytic networks studied in \cite{htk1}.
% ---------- PKv08 START
In the model of Sol\'e and Manrubia \cite{solemanrubia}
 only linear interactions are allowed (i.e. $\vert j \vert=1$ in Eq.(\ref{EvoGen})) and new species are created not through endogenous recombinations, but by an explicit mutation mechanism.
As discussed in \cite{tkh09}, $f_i(\vec\sigma(t))$ in our 
combined stochastic
model plays the identical role as the randomly assigned fitness values in the Bak-Sneppen model \cite{Bak93}.
To recover the $NK$-model \cite{kauffman_NK} as a special case associate each species with a bit-string.
A random fitness value is then assigned to each species' bit-string, in some variants of the model also in dependence of a given number of bits of other species' strings.
However, fitness in our framework is a topological property of the entire system plus the set of abundant species, whereas in $NK$-models fitness is basically a mapping of random numbers to bit-strings.
% ---------- PKv08 END

We find that the model of constructive and destructive interactions reproduces stylized facts of man-made (economies) and natural evolutionary systems (metabolic networks, macro-ecology) across different orders of magnitude. We belief this adds empirical substance to our claim that we have identified a crucial and ubiquitous building block of evolutionary systems with recombinatory, non-linear interactions within a simple binary framework.    
The model systematically expands on the idea that the concept of fitness is an {\em a posteriori} concept. Fitness in the traditional sense can of course be reconstructed for every timestep in our model.  It is nothing but the co-evolving network of rates  of the actually active (productive) processes at a given time, see \cite{tkh09} for more details. 
It becomes clear that fitness can not be used as concept with much predictive value, even if `Darwin's Demon' knowing all mutual influences at a given time would exist.
 The proposed model is free of `Darwin's Demon'.

This work was supported in part by the Austrian Science Fund, FWF P19132.

\end{document}